# Temperature- and Magnetic-Field-Tuning of Magnetic Phases in Multiferroic $NdFe_3(BO_3)_4$

Christie S. Nelson[*]

*Photon Sciences Directorate, Brookhaven National Laboratory, Upton, NY 11973, USA*

Leonard N. Bezmaternykh and Irina A. Gudim

*L.V. Kirensky Institute of Physics, Siberian Branch of RAS, Krasnoyarsk, 660036, Russia*

We report the low-temperature coexistence in $NdFe_3(BO_3)_4$ of an incommensurate magnetic phase with a strained commensurate magnetic phase that is primarily at the surface of the crystal. Increasing the temperature or magnetic field decreases the incommensurability and stabilizes the commensurate magnetic phase above $T_{ic} \approx 14$ K or $H_{ic} = 0.9$ T. A comparison to published studies indicates the onset of longitudinal magnetostriction and electric polarization at the magnetic-field-induced transition, which may arise due to a basal plane spin-flop and canting of moments along the field direction.

[*]Email: csnelson@bnl.gov

# I. INTRODUCTION

Interest in the rare earth ferroborates, $RFe_3(BO_3)_4$, focuses on their potentially useful optical properties,[1,2] as well as the multiferroic behavior[3,4] exhibited by some of these materials. In $NdFe_3(BO_3)_4$ in particular, a sizable ~400 $\mu C/m^2$ magnetic-field-induced polarization in the antiferromagnetic phase[4] has stimulated interest, and detailed studies of the zero field magnetic structure have been carried out via neutron scattering.[5,6] Simultaneous ordering of the Nd and Fe moments at $T_N = 30.5$ K has been reported,[5] with a commensurate magnetic structure for temperatures between ~13.5 K and $T_N$, and an incommensurate magnetic structure below ~13.5 K.[6] The Nd and Fe moments are observed to be in the basal plane in both phases.[6]

One question that previous studies of $NdFe_3(BO_3)_4$ have not addressed concerns its magnetic structure in an applied magnetic field. Given the interesting magnetic-field-induced properties exhibited by $NdFe_3(BO_3)_4$, such as electric polarization as well as longitudinal magnetostriction,[4] information about the magnetic structure is important for providing insight into their origin. In this paper we attempt to shed light on this issue by reporting the behavior of the magnetic phases in $NdFe_3(BO_3)_4$ at the onset of multiferroicity. We observe the low-temperature coexistence of incommensurate and strain-induced commensurate magnetic phases, which indicates sensitivity of the magnetic structure to the lattice degree of freedom, and a field-induced incommensurate-to-commensurate transition at $H_{ic} = 0.9$ T.

# II. EXPERIMENTS AND DISCUSSION

Single crystals of $NdFe_3(BO_3)_4$ were grown by the flux method,[7] and the mosaic spread as measured at the (003) Bragg peak was 0.03°. X-ray scattering measurements were carried out on wiggler beamline X21 at the National Synchrotron Light Source at Brookhaven National Laboratory. This beamline employs a cryo-cooled Si(111) double crystal monochromator and focusing mirrors to deliver beam to tandem hutches with multiple endstations, including a high-field magnet diffractometer and a 4-circle diffractometer.

The high-field magnet diffractometer uses a horizontal scattering geometry, with magnetic field applied by a split-coil, superconducting magnet in the vertical direction, while the 4-circle uses a vertical scattering geometry and a closed-cycle displex for sample cooling. The base temperatures are 1.8 and ~7 K for the magnet and displex, respectively. In the high-field magnet, the $NdFe_3(BO_3)_4$ crystal was oriented such that the magnetic field was applied in the basal plane, along the crystallographic *a* axis. With both instruments, structural and magnetic peaks were measured along [00*l*], and LiF(200) was used as an analyzer for background suppression. Resonant x-ray scattering was carried out at the peak of the resonance in the vicinity of the Nd $L_2$ edge. Nonresonant x-ray scattering, which is primarily sensitive to the Fe ions given the 5:1 ratio of spin moment per unit cell of $Fe^{3+}$ to $Nd^{3+}$ ions, was measured at an energy of 6.65 keV.

With the sample cooled to the base temperature of the magnet or displex, two types of peaks, as shown in Figure 1, are observed due to magnetic ordering: at (*hkl*) ± (0 0 1.5) and (*hkl*) ± (0 0 1.5 ± δ), where (*hkl*) is a structural Bragg peak. The incommensurability, δ, at T = 1.8 K is 0.0065 c*, and decreases as the incommensurate-to-commensurate transition temperature, $T_{ic}$, is approached. Although commensurate peaks below $T_{ic}$ were not observed with neutrons,[6] the commensurate peaks observed with x-rays are at the fundamental energy and exhibit a strong resonant enhancement, as shown in Figure 2. Our results indicate that for T < $T_{ic}$, both commensurate and incommensurate magnetic phases are present in $NdFe_3(BO_3)_4$.

To learn more about the coexistent magnetic phases for T < $T_{ic}$, the q-dependences of the commensurate and incommensurate magnetic peaks were measured at resonance. As can be seen in Figure 3(a), none of the magnetic peaks in this temperature regime are resolution-limited. At T = 2 K, the incommensurate peaks at (0 0 1.5 ± δ) and (0 0 4.5 ± δ) are of similar width, with a c-axis correlation length of $\xi_c = 430 \pm 20$ Å.[8] In contrast, the widths of the commensurate peaks at (0 0 1.5) and (0 0 4.5) are q-dependent: the (0 0 4.5) peak is ~50% broader than the (0 0 1.5) peak. Such a q-dependence with broadening as a function of increasing q is indicative of the presence of strain.

A comparison of the intensities of the magnetic peaks provides additional information about the strained commensurate magnetic phase. The integrated intensities are plotted as a function of temperature in Figure 3(b). At T = 2 K, the commensurate (0 0 1.5) peak is more than four times stronger than the commensurate (0 0 4.5) peak, which is unexpected for the easy-plane magnetic phase. This can be seen by calculating the geometrical dependence of the resonant scattering cross-section for incident π polarization and dipole, 2p → 5d, transitions, which is proportional to $\cos^2(\theta)$ + $\sin^2(2\theta)\cos^2(2\theta_a)$, where θ is 10.5° and 33.1° for the (0 0 1.5) and (0 0 4.5) magnetic peaks, respectively, and $2\theta_a$ is the scattering angle of the analyzer.[9] Based on the geometrical dependence, the two commensurate peaks should be nearly identical in intensity. The unexpected q-dependence of the intensity that we observe therefore suggests a surface origin of the strained commensurate magnetic phase, which would also be consistent with neutron results given the bulk sensitivity of that probe.

The data in Figure 3 also indicate the evolution of the incommensurate and strained commensurate magnetic phases as the incommensurate-to-commensurate transition is approached. The incommensurate peaks near (0 0 1.5) and (0 0 4.5) decrease in intensity at similar rates, while the two commensurate peaks are more weakly temperature dependent. From neutron scattering measurements by Fischer and co-workers,[5] the Nd moment is expected to decrease by a factor of ~2 between T = 2 and 15 K, which can explain most of the decrease in intensity observed at the incommensurate positions. The additional intensity decrease and the weak temperature dependence of the commensurate peak intensities suggests growth of the strained, surface commensurate phase at the expense of the incommensurate phase as the transition is approached, however with no change in the correlation length. Long-range order of the commensurate magnetic phase does set in at $T_{ic} \approx 14$ K, above which both the (0 0 1.5) and (0 0 4.5) peaks are resolution-limited.

The temperature dependences of the commensurate and incommensurate peaks were also measured off resonance, and the integrated intensities are shown in Figure 4. Below $T_{ic}$, the decrease in intensity of the incommensurate peaks and the weak temperature dependence of the commensurate peak are similar as to what is observed at resonance. This indicates a close coupling of the Fe and Nd subsystems in the two phases. Above $T_{ic}$ the decrease in intensity reflects the decrease in the ordered Fe moment as $T_N$ is approached.

The low-temperature phase coexistence of commensurate and incommensurate magnetic phases was also studied in an applied magnetic field, and reciprocal space scans measured at resonance at T = 1.8 K are shown in Figure 5. Negligible changes are observed in magnetic field up to H = 0.6 T, while the incommensurability decreases and there is a transfer in scattering intensity from the incommensurate to commensurate peaks at H = 0.8 T. The transfer is complete and only a commensurate peak is observed at $H_{ic} = 0.9$ T.

The magnetic-field-induced, incommensurate-to-commensurate magnetic transition can be correlated with changes in longitudinal magnetostriction and electric polarization, as reported by Zvezdin et al.[4] In this work, a critical field of ~1 T applied along *a* was observed below T = 25 K, and was attributed to a spin-flop transition in domains with moments along *a* in zero field.[4] Such a basal plane flop of the moments would result in an increase in scattering intensity, as the moments would lie in the scattering plane. In fact just such an increase is observed— the integrated intensities of the magnetic peaks measured near (0 0 4.5), shown in Figure 5, increase by ~25% between H = 0 and 0.9 T. Assuming an initial equal domain population with moments along the three second-order axes in the basal plane, an intensity increase of ~43% ($\frac{1}{2}[\cos^2(\theta) + \sin^2(2\theta)\cos2(2\theta_a)] \rightarrow \cos^2(\theta)$) is expected, which suggests some canting of the moments along **H**. We note that a correlation between magnetic-field-induced electric polarization, an incommensurate-to-commensurate magnetic transition, and basal plane moments with canting has also been reported for multiferroic $GdFe_3(BO_3)_4$.[10]

## III. CONCLUSION

In conclusion, we report the low-temperature coexistence of incommensurate and commensurate magnetic phases in $NdFe_3(BO_3)_4$, with the commensurate magnetic phase arising due to strain near the surface of the sample. Such sensitivity of the ground state incommensurate magnetic phase to perturbations is also demonstrated by our observation that a magnetic field of 0.9 T applied along the crystallographic *a* axis tips the balance between the two magnetic phases to stabilize the commensurate magnetic phase, and this correlates with the onset of longitudinal magnetostriction and electric polarization.[4] In addition to the commensurate magnetic structure, two additional characteristics that may be related to the multiferroicity in $NdFe_3(BO_3)_4$ are a basal plane spin-flop and canting along **H**.

## ACKNOWLEDGMENT

We thank S. LaMarra for his assistance with the magnet at beamline X21. Use of the National Synchrotron Light Source, Brookhaven National Laboratory, was supported by the U.S. Department of Energy, Office of Science, Office of Basic Energy Sciences, under Contract No. DE-AC02-98CH10886.

Figure Captions.

Fig. 1. Reciprocal space scans along [00l] measured at resonance as a function of temperature.

Fig. 2. Scattering intensity of the commensurate (0 0 1.5) peak (■), at T = 7 K, and sample fluorescence (−) as a function of incident photon energy.

Fig. 3. Widths (a) and integrated intensities (b), measured at resonance, of commensurate and incommensurate magnetic peaks. The integrated intensities are normalized to the (003) structural Bragg peak integrated intensities, and those near (0 0 1.5) have been multiplied by 1.7 to correct for beam leakage at small incidence angles.

Fig. 4. Integrated intensities, measured off resonance, of commensurate and incommensurate magnetic peaks.

Fig. 5. Reciprocal space scans along [00$l$] measured at resonance at T = 1.8 K as a function of applied magnetic field.

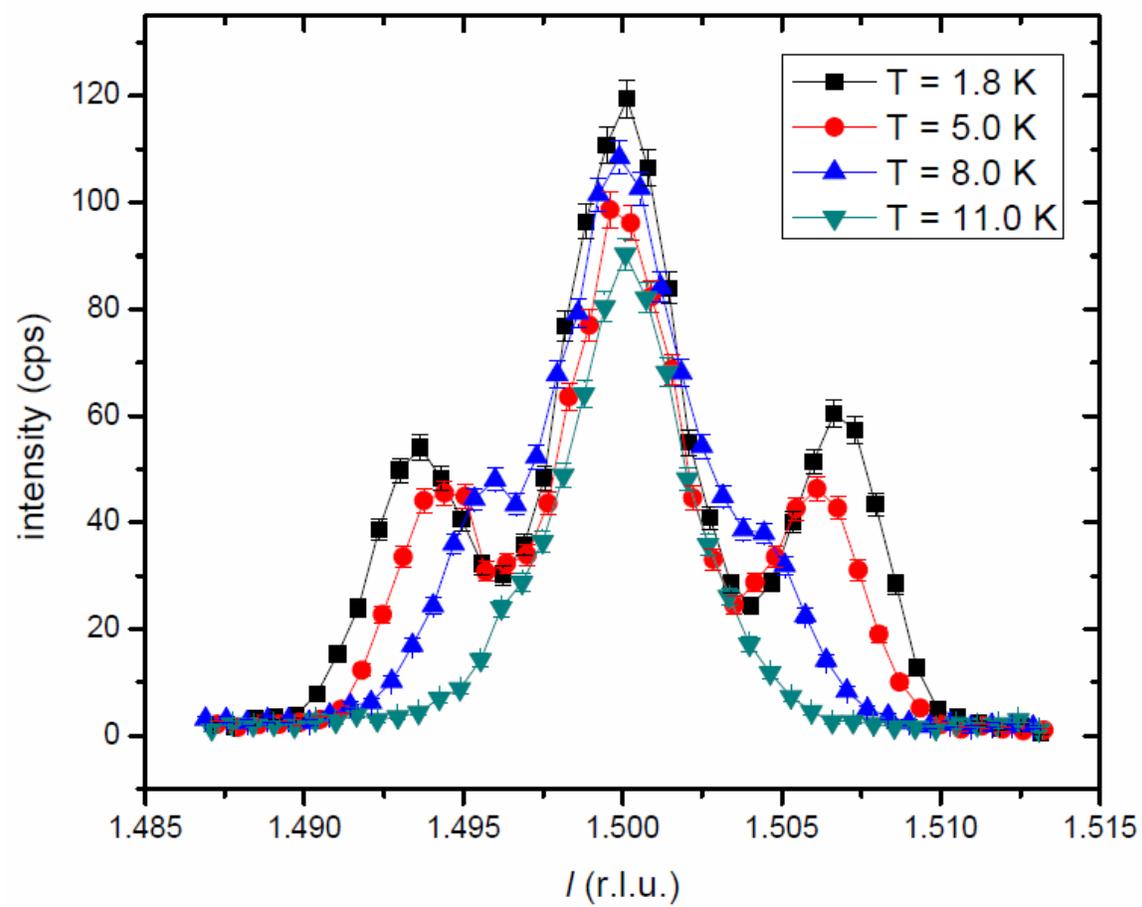

Fig. 1

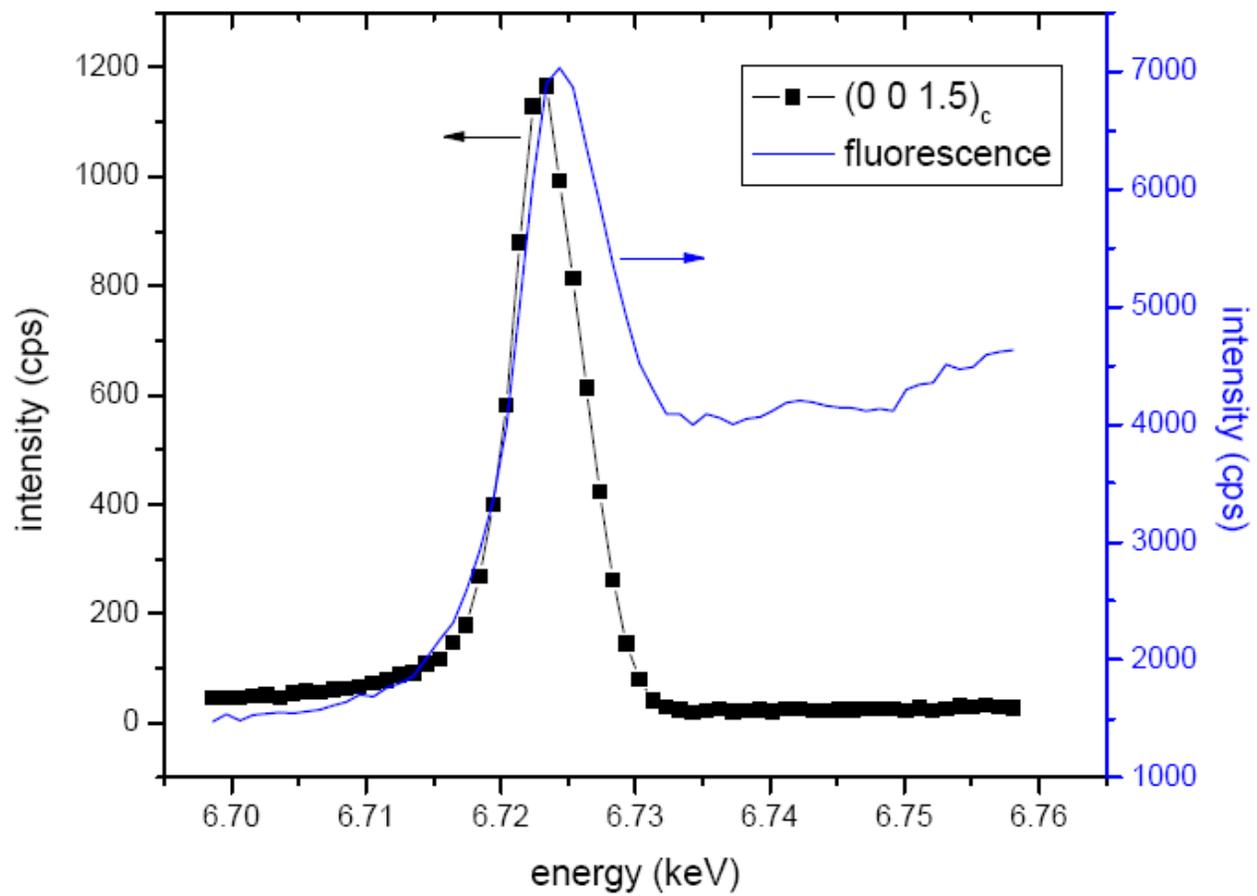

Fig. 2

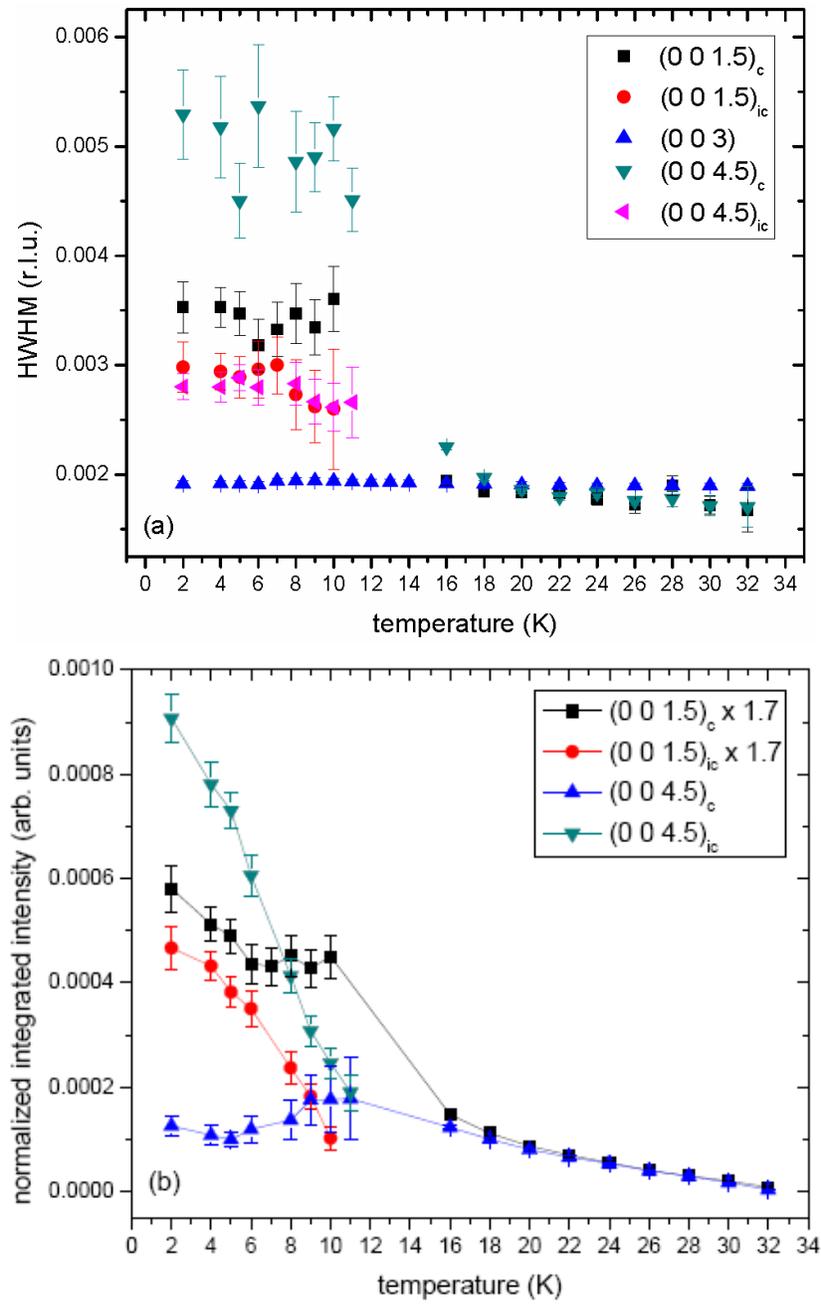

Fig. 3

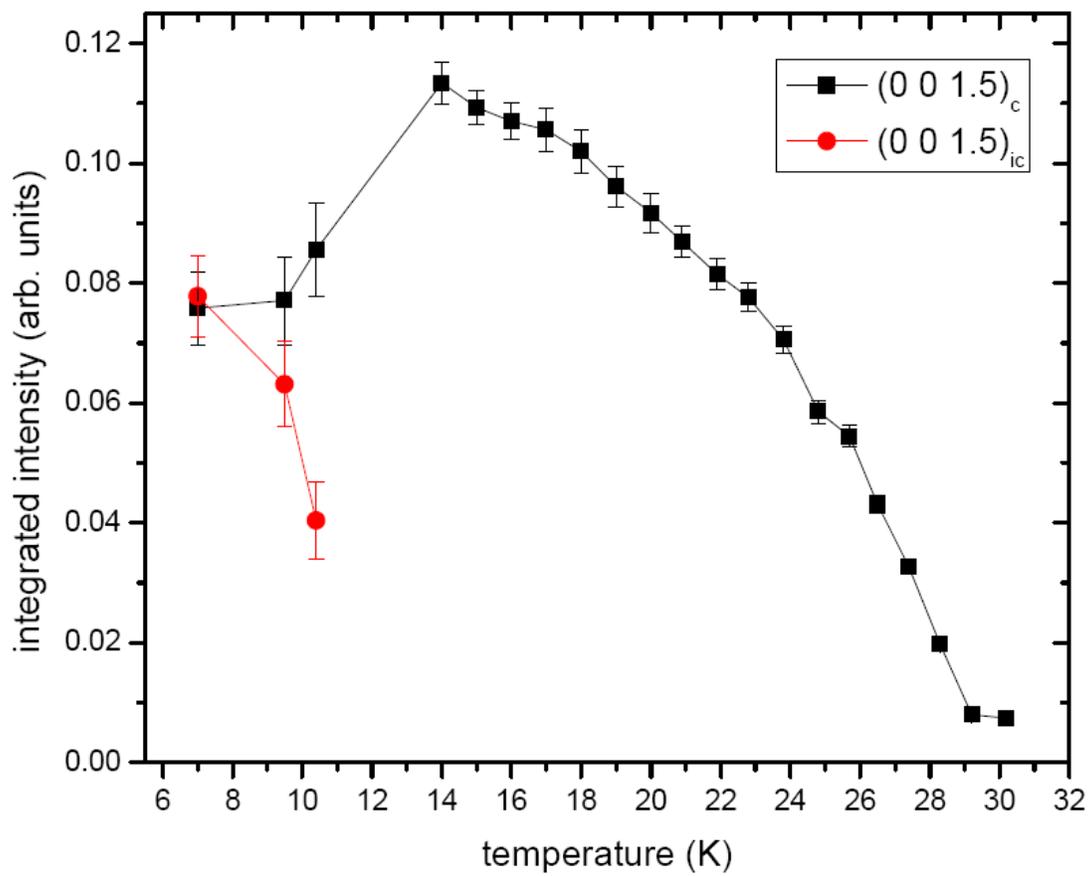

Fig. 4

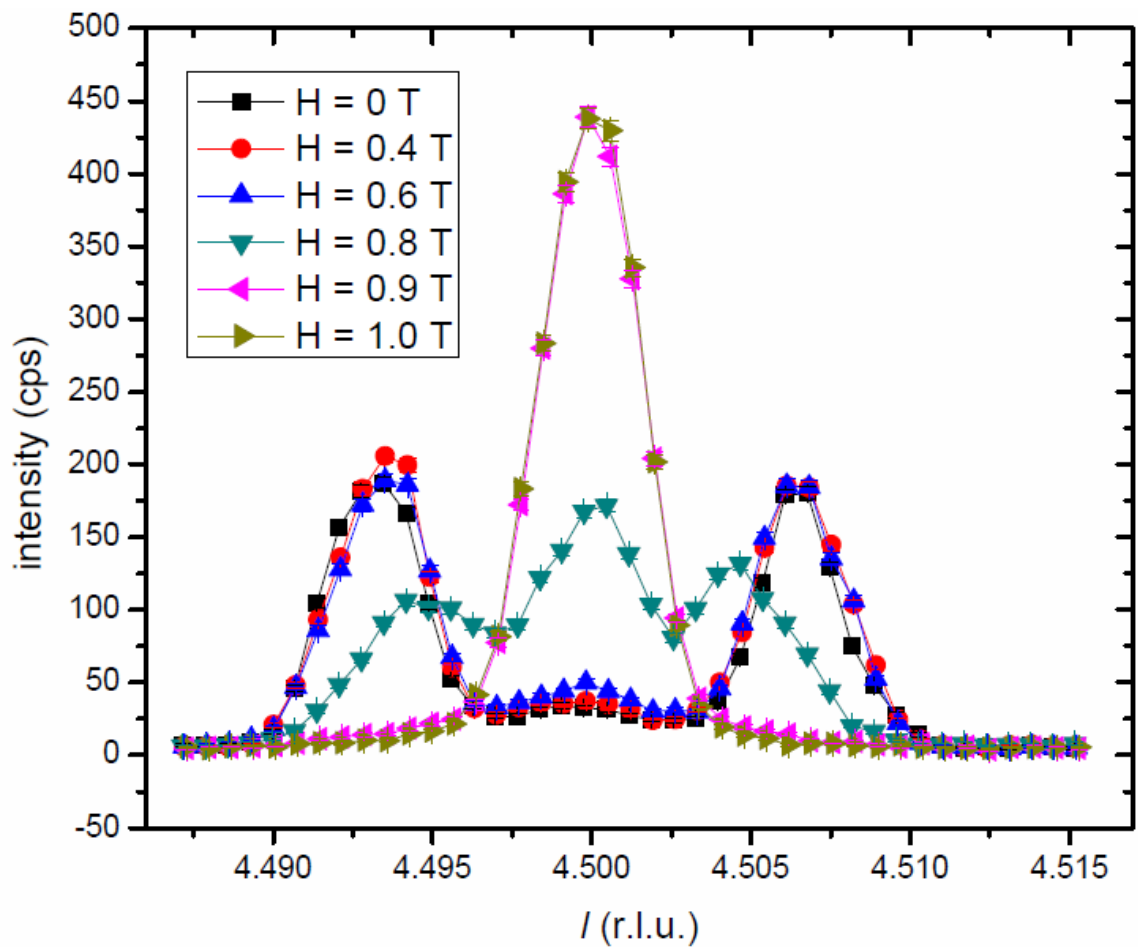

Fig. 5